%% file: main.tex
\documentclass{article}
\usepackage[preprint]{spconf}

\copyrightnotice{\copyright\ IEEE 2023}
\toappear{To appear in {\it Proc.\ ICASSP2023,
    June 04-10, 2023, Rhodes Island, Greece}}

\usepackage[utf8]{inputenc}
\usepackage[english]{babel}
\usepackage{booktabs}
\usepackage{etoolbox}
\usepackage{subcaption}
\usepackage{color, colortbl}
\usepackage{longtable}
\usepackage{xcolor}
\usepackage{soul}
\usepackage{comment}
\usepackage{url}
\usepackage{balance}
\usepackage{graphicx}
\usepackage{siunitx}
\usepackage{balance}
\usepackage[capitalise,noabbrev]{cleveref}

\title{Heart Rate Extraction from Abdominal Audio Signals}

\name{%
\begin{tabular}{@{}c@{}}
Jake Stuchbury-Wass$^\text{\normalfont{1}}$,
Erika Bondareva$^\text{\normalfont{1}}$, 
Kayla-Jade Butkow$^\text{\normalfont{1}}$,  
Sanja Šćepanović$^\text{\normalfont{2}}$,\\
\textit{Zoran Radivojevic$^\text{\normalfont{2}}$,
Cecilia Mascolo$^\text{\normalfont{1}}$}
\end{tabular}}

\address{$^\text{1}$University of Cambridge, UK;\\ $^\text{2}$Nokia Bell Labs, Cambridge, UK}
\begin{document}

\ninept
\maketitle

\input{sections/abstract}

\section{Introduction}

\input{sections/intro}

\section{Methodology}

\input{sections/method}

\section{Results and Discussion}
\label{sec:results}
\input{sections/results}

\section{Conclusions and Future Work}

\input{sections/conclusion}

\section{Acknowledgments}
This work is supported by ERC through Project 833296 (EAR), EPSRC grant EP/L015889/1 and EP/S023046/1 for the EPSRC CDT in Sensor Technologies and Applications, EVIDEN/New Atos Life Sciences Centre of Excellence, and Nokia Bell Labs.

\bibliographystyle{IEEEbib}
\balance
\bibliography{references}

\end{document}

%% file: sections/abstract.tex
\begin{abstract}
Abdominal sounds (ABS) have been traditionally used for assessing gastrointestinal (GI) disorders. However, the assessment requires a trained medical professional to perform multiple abdominal auscultation sessions, which is resource-intense and may fail to provide an accurate picture of patients' continuous GI wellbeing. This has generated a technological interest in developing wearables for continuous capture of ABS, which enables a fuller picture of patient's GI status to be obtained at reduced cost. This paper seeks to evaluate the feasibility of extracting heart rate (HR) from such ABS monitoring devices. The collection of HR directly from these devices would enable gathering vital signs alongside GI data without the need for additional wearable devices, providing further cost benefits and improving general usability. We utilised a dataset containing 104 hours of ABS audio, collected from the abdomen using an e-stethoscope, and electrocardiogram as ground truth. Our evaluation shows for the first time that we can successfully extract HR from audio collected from a wearable on the abdomen. As heart sounds collected from the abdomen suffer from significant noise from GI and respiratory tracts, we leverage wavelet denoising for improved heart beat detection. The mean absolute error of the algorithm for average HR  is 3.4\,BPM with mean directional error of -1.2\,BPM over the whole dataset. A comparison to photoplethysmography-based wearable HR sensors shows that our approach exhibits comparable accuracy to consumer wrist-worn wearables for average and instantaneous heart rate.
\end{abstract}

%% file: sections/intro.tex
Abdominal sounds (ABS) have traditionally been used by medical professionals to monitor gastrointestinal (GI) activity~\cite{cannon_auscultation_1905}.
In recent years, ABS collected with an e-stethoscope in a clinical setting with applied digital signal processing techniques have been used to uncover features specific to certain GI disorders~\cite{ranta2005complete}. Modern methods such as machine learning classifiers have also been deployed on ABS data to diagnose GI disorders~\cite{allwood2018advances}. However, abdominal auscultation has to be performed by specially trained doctors, and can only be done at discrete intervals for a relatively short time duration due to its resource-intensity. To enable continuous monitoring of ABS, interest in abdominal wearable devices has soared~\cite{wells2022wearable, wang2019flexible}. These wearable devices collect sounds audible from the abdomen, with the primary goal of capturing GI sounds. Examples include early meal onset detection~\cite{kumar2019pilot} and user's stress level detection~\cite{bondareva2022stress}. However, because of the close proximity of the abdomen to the thoracic cavity, heart and respiratory sounds are inadvertently captured, too. 

Heart sounds are clinically extremely valuable as they can be used to diagnose heart-based pathologies, as well as for determining heart rate and related vital signs. Heart activity is measured in the clinical setting using phonocardiogram (PCG), where sound is captured, or using electrocardiogram (ECG), which captures heart's electrical activity. In wearables, the most prevalent sensing modality for continuous heart rate (HR) monitoring is photoplethysmography (PPG), which uses the changes in light reflectance due to expansion and contraction of the blood vessels~\cite{laukkanen1998heart}. Two key vital signs that are typically used for heart activity evaluation are average and instantaneous heart rate. HR is an early indicator of cardiovascular disease (CVD), which is the leading cause of  death globally with an estimated 17 million deaths from CVDs per year~\cite{who_global_2012}.
Abnormalities in average HR are linked to mortality~\cite{ljunggren2016association}, while instant HR has been shown to have applications in seizure monitoring~\cite{smith1989profiles}. In addition, instant HR can be used to calculate heart rate variability (HRV), a metric associated with mortality~\cite{fang_heart_2020}.

\begin{figure}[t]
\centering
	\includegraphics[width=8.6cm]{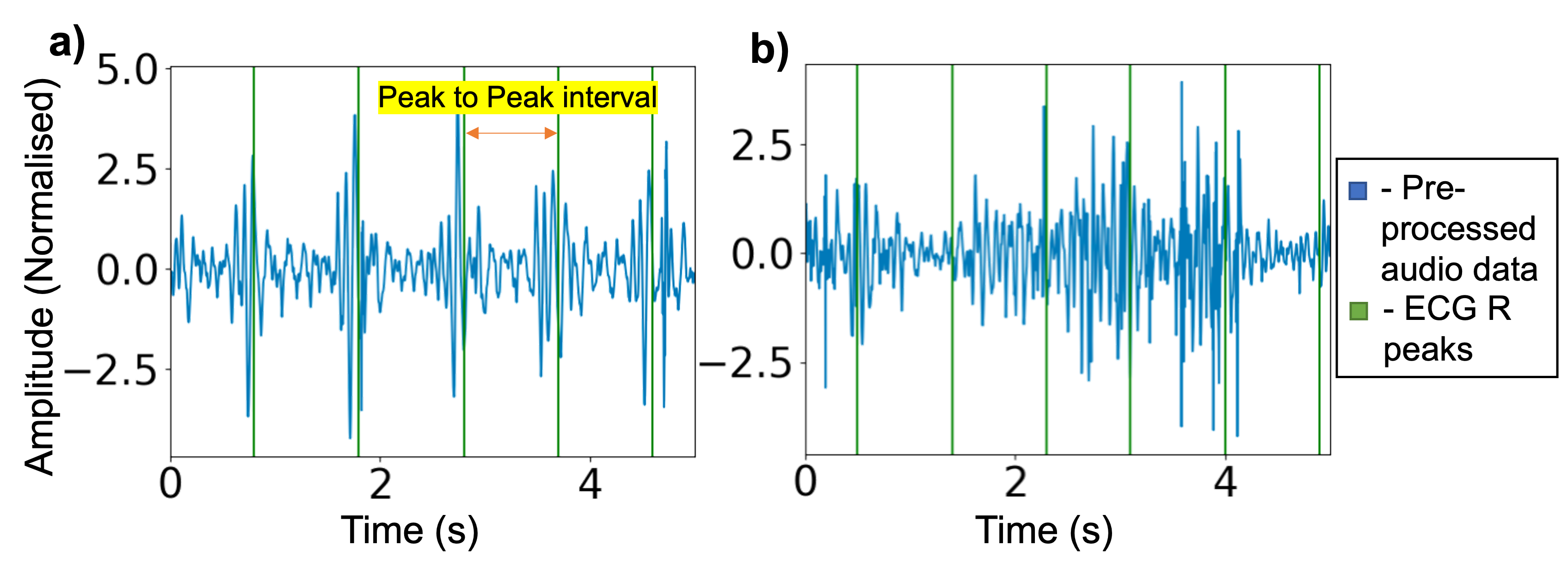}
	\caption{
		Example of (\textbf{a}) clean and (\textbf{b}) noisy audio recorded from the abdomen. The audio data is annotated with ECG R peaks as well as the peak to peak interval used to calculate heart rate.}
	\label{fig:abs_data}
\end{figure}

This work explores the feasibility of vital signs extraction from audio collected using a wearable device, leveraging PCG by using heart sounds that travel through the thorax to the abdomen. Unlike in traditional chest-based PCG where clean heart sounds are easier to obtain, extracting vital signs from ABS brings multiple challenges. Firstly, the position of the microphone on the lower abdomen ensures high quality GI sounds, but the signal to noise ratio (SNR) of heart sounds suffers significantly. Secondly, heart sounds collected at the abdomen are obfuscated by abdominal and breathing sounds with high SNR, making heart sound analysis challenging. Thirdly, sounds created by movement of the user are captured with the microphone adding additional complexity to the signal. \cref{fig:abs_data} shows an example of audio recorded at the abdomen where heart sounds are clear and where the heart sounds are obscured by noise.

This paper presents, for the first time, a feasibility study into determining vital signs from ABS, a key step in achieving a minimally-obtrusive, suitable for daily wear abdominal health wearable. We study the potential of extracting average heart rate (aHR) and instantaneous heart rate (iHR) from ABS captured using a custom-built e-stethoscope embedded in a wearable belt. Audio collected using a microphone on the lower abdomen goes through several stages of signal processing including lowpass filtering, wavelet denoising, and peak detection to estimate average and instantaneous HR, as well as post-processing for outliers. We compared our results to those obtained using a ground truth ECG chest strap. The approach demonstrates mean absolute percentage errors of 4.8\% for average HR and 8.9\% for instantaneous heart rate from the abdominal audio.

%% file: sections/method.tex
\subsection{Dataset}
The dataset~\cite{bondareva2022stress} used for this work consists of audio collected from the abdomen with a custom-built e-stethoscope embedded in a stretchable belt, as well ECG ground truth data captured with a Polar H9 chest strap. The data collection was approved by the ethics committee of the Department of Computer Science and Technology at the University of Cambridge.
The participants in this study were seated at rest, therefore their heart rate can be expected to be in the standard  50-90\,BPM range~\cite{nanchen_resting_2018}. Overall, 104 hours of audio and ECG data was collected from 7 participants across 10 days.
The data were collected remotely during the pandemic, leading to variability in adherence to the data collection instructions across participants.
To account for this, a data quality check was performed to remove fragmented or excessively noisy audio or ECG samples.

\subsection{Algorithm Overview}
Vital sign extraction from sounds captured on the abdomen is non-trivial. The SNR of the heart sounds is very poor due to the positioning of the e-stethoscope on the abdomen and due to the interference of other sounds in the same frequency range, such as the sounds from GI  and respiratory tracts. This section presents our solution to overcoming these challenges and performing, for the first time, extraction of vital signs from audio collected on the abdomen.

\begin{figure}[h]
    \centering
	\includegraphics[width=8.6cm]{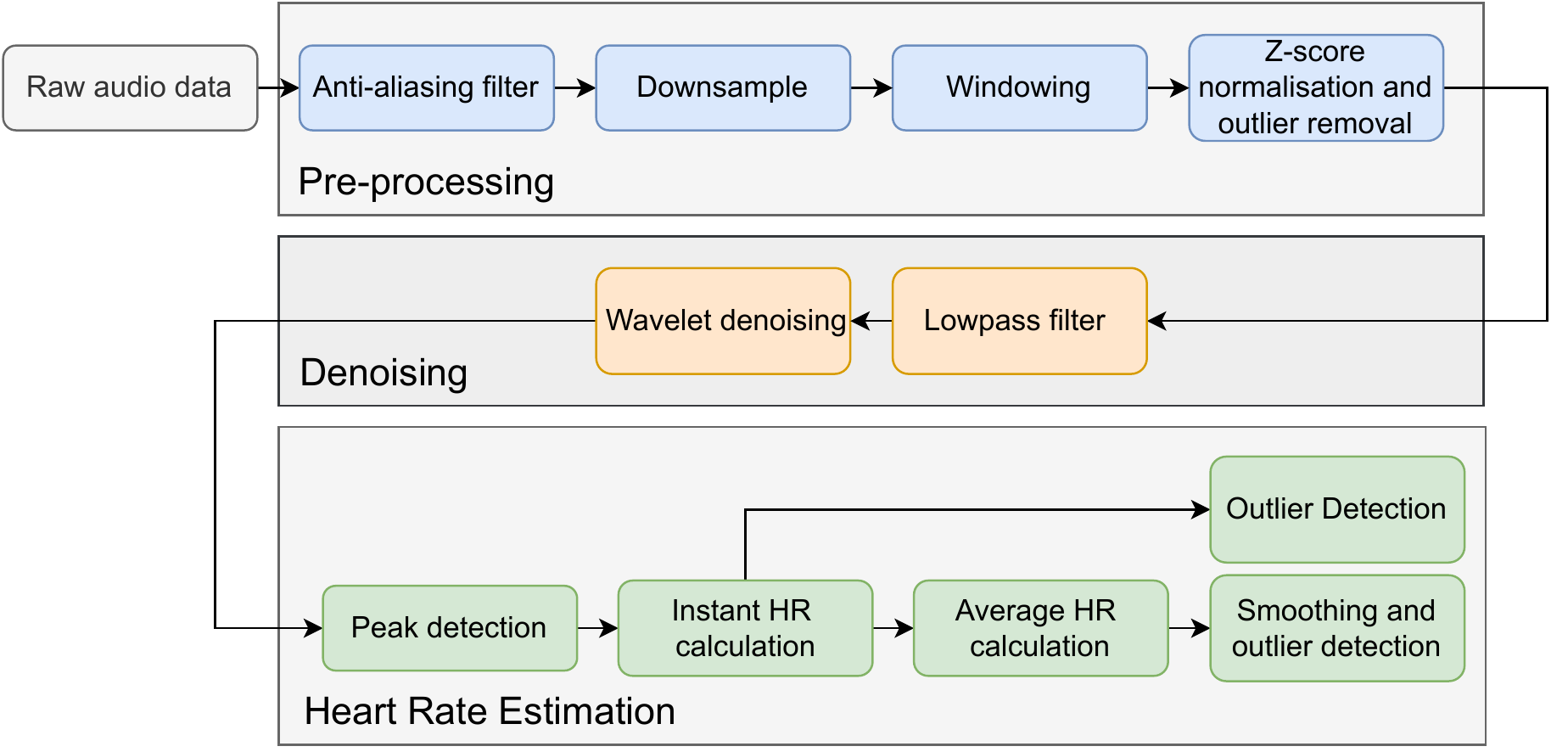}
	\caption{
		Block diagram of the algorithm developed for vital sign extraction from audio recorded at the abdomen.
	}
	\label{fig:blockdiagram}
\end{figure}

The general structure of our designed algorithm for vital sign extraction from audio recorded at the abdomen is shown in \cref{fig:blockdiagram}. The following subsections describe each block in detail.

\subsubsection{Pre-processing}

We first apply an anti-aliasing low-pass filter with 2\,kHz cutoff and then downsample the signal to 4\,kHz. Secondly, we split the data into 10\,s windows, each with a 2\,s overlap with the previous and the following window. A 10\,s window is used in the literature~\cite{li2007robust} and should contain roughly 10 heart sounds, which we deemed sufficient for averaging (for aHR extraction), while still being short enough to capture short-term temporal changes in heart data. We then apply z-score normalisation to the data. This allows us to identify outliers in the audio, which predominantly occur because of motion artefacts. After z-score normalisation, if the window contains samples lying more than 10 standard deviations from the mean, we discard the window. By doing this, we retain only windows that are free from severe motion artefacts to improve robustness of HR estimation.

\subsubsection{Signal Denoising}
After pre-processing, windows with severe motion artefacts are removed. However, some environmental noise, as well as undesired sounds from GI and respiratory tracts are still present within the signal. To reduce the impact of this noise, we apply denoising to the retained windows. 

Specifically, we use a 5th order Butterworth lowpass filter with a 200\,Hz cutoff frequency. 
This filter attenuates high-frequency information that is related to external noise, and retains information related to internal body sounds. Since internal body sounds lie in a similar frequency range,  band-pass filtering tends to be ineffective for separating heart sounds from noise coming from GI and respiratory tracts. To isolate the heart sounds, we apply a wavelet-based filtering technique. We apply the discrete wavelet transform (DWT) to each window and modify the coefficients with soft thresholding~\cite{donoho1994ideal}. We then perform the inverse DWT to obtain the cleaned heart sounds. We use the Coiflet 4 wavelet with the 5$^\text{th}$ decomposition level as these were empirically determined to have the best performance. When applying the threshold, we use the \textit{sqtwolog} method with soft thresholding~\cite{donoho1994ideal}. This method allows us to isolate and remove coefficients with high variance from the mother wavelet which has a low variance from the heart sounds. This is detailed in \cref{eq:sqtwolog}, where $\sigma_j$ is the mean absolute deviation of the wavelet coefficients, $N_j$ is the length of the signal, and $j$ is the decomposition level.

\begin{equation}
    th_j = \sigma_j \sqrt{2log(N_j)}
    \label{eq:sqtwolog}
\end{equation}

\subsubsection{Heart Rate Estimation}\label{sec:vitals}

After pre-processing and denoising, we extract vital signs from the cleaned audio signals. In the denoised signal, peaks correspond to heart beats (as shown in \cref{fig:abs_data}). We thus detect peaks directly from the cleaned signal, and calculate the timings between consecutive peaks. We use a minimum distance between peaks of 0.65\,s and minimum peak height of 1.2. The peak distance is chosen to allow for a resting heart rate of 92\,BPM, which is over the maximum HR typical for a resting state, as stated by~\cite{nanchen_resting_2018}, and the height is empirically determined by examining the amplitudes of the cleaned data. 

The instantaneous HR (iHR) is calculated using the time difference between two consecutive peaks, the P-P time (\cref{eq:ihr}). We perform a final check on the predicted heart beats to ensure they lie within a reasonable heart rate range. If detected heart sounds are too far apart, then this P-P interval is discarded and the next interval is used, the threshold for this rejection is a heart rate under 45\,BPM. (i.e. outside the human resting HR range~\cite{nanchen_resting_2018}). The average HR (aHR) is then calculated by averaging the iHR samples over the window. Due to the overlapping windows, the last 2 seconds of each window are not included in the aHR calculation to avoid double counting their contribution.

\begin{equation}
    \text{iHR}=\frac{60}{\text{P-P time}}
    \label{eq:ihr}
\end{equation}

The aHR is post-processed to remove outliers and smooth the results. We use a moving standard deviation window, with a threshold of two standard deviations,  to detect and replace outliers. Finally, we smooth the signal using a moving average filter across the windowed results. 

%% file: sections/results.tex
\subsection{Metrics}

The metrics used to assess system performance are Mean Directional Error (MDE) (Equation~\ref{eq:me}), Mean Absolute Error (MAE) (Equation~\ref{eq:mae}) and Mean Absolute Percentage Error (MAPE) (Equation~\ref{eq:mape}). N is the number of samples, $\text{HR}_{\text{Audio}}$ represents samples generated from audio data, and $\text{HR}_{\text{ECG}}$ represents samples from the ground truth ECG data. These metrics were used for evaluation of both aHR and iHR.

\begin{equation}
    \text{MDE} = \frac{\sum \text{HR}_{\text{Audio}}-\text{HR}_{\text{ECG}}}{\text{N}}
    \label{eq:me}
\end{equation}

\begin{equation}
    \text{MAE} = \frac{\sum abs(\text{HR}_{\text{Audio}}-\text{HR}_{\text{ECG}})}{\text{N}}
    \label{eq:mae}
\end{equation}

\begin{equation}
    \text{MAPE} = 100\cdot\frac{\sum \frac{abs(\text{HR}_{\text{Audio}}-\text{HR}_{\text{ECG}})}{\text{HR}_{\text{ECG}}}}{\text{N}}
   \label{eq:mape}
\end{equation}

\subsection{Overall Performance}

We assessed the feasibility of monitoring heart rate via abdominal sounds by evaluating our algorithm's accuracy in estimating aHR and iHR in participants at rest, as defined in \cref{sec:vitals}. \cref{tab:results} shows the MDE and MAE across the entire dataset for iHR and aHR. The results show very good accuracy for both the iHR and aHR, with average errors under 10\%, the threshold for medical-grade accuracy under resting conditions~\cite{PhysicalActivityMonitoringForHeartRateANSI}. Worth noting, errors for iHR are consistently higher than for aHR, but this is expected since aHR averages out multiple iHRs within a window.

\begin{table}[ht]
\centering
\caption{Results of iHR and aHR across the whole dataset}
\begin{tabular}{@{}llll@{}}
\toprule
    & MDE (BPM) & MAE (BPM) & MAPE (\%) \\ \midrule
iHR  & -0.18 & 6.4 & 8.9 \\
aHR  & -1.2 & 3.4 & 4.8 \\ \bottomrule
\end{tabular}
\label{tab:results}
\end{table}

\begin{figure}[ht]
    \centering
    \includegraphics[width=4.0cm]{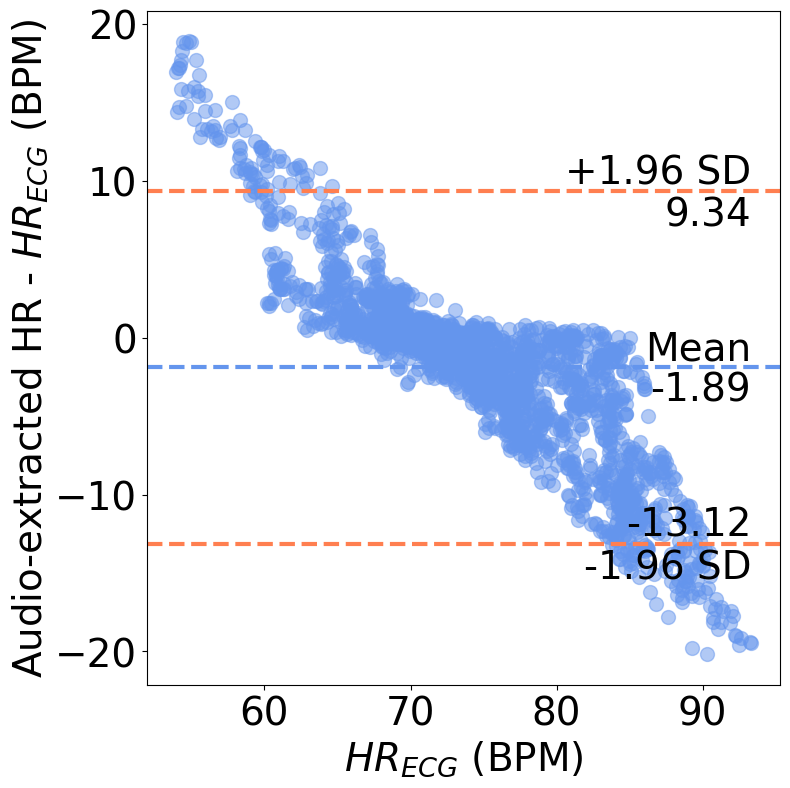}
    \includegraphics[width=4.0cm]{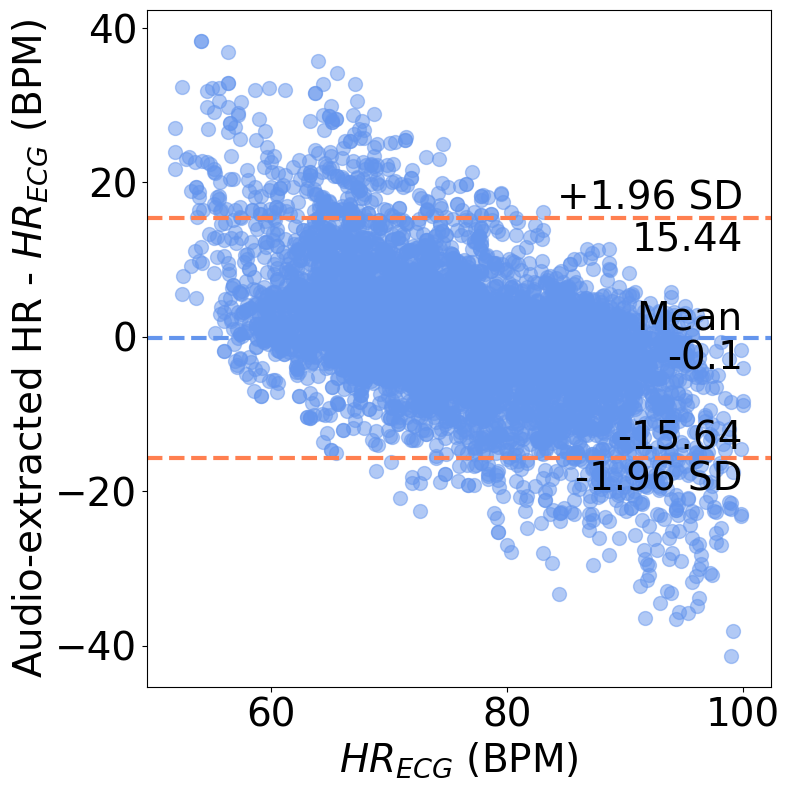}
    \caption{Bland-Altman diagrams for aHR (left) and iHR (right) for the dataset (randomised subset of data shown for visibility).}
    \label{fig:blandaltman}
\end{figure}

\cref{fig:blandaltman} provides a modified Bland Altman (BA) plot of HR estimation for aHR and iHR (left and right respectively).
BA plots indicate the bias between the measured and GT value for each GT value. BA plots are used clinically to assess the level of agreement between two measurement methods~\cite{Giavarina2015UnderstandingAnalysis}.  The data shown in the figure is a subset of the full results set, which was randomly sampled from each participant for ease of visualisation. The BA plots indicate a low error and small limits of detection for both aHR and iHR. However, for aHR, a very clear trend exists whereby the system overestimates heart rates under 60\,BPM and underestimates heart rates over 85\,BPM. While this trend also exists for iHR, it is less pronounced than for aHR. This is due to the parameters of the peak detection algorithm which enforce a maximum heart rate of 92\,BPM. While this design decision is consistent with literature, it biases the detected heart rates toward lower values. This effect can also be seen in the peaks and troughs of \cref{fig:track}. When choosing the techniques used in the system, we favoured lightweight, lower complexity algorithms that can be implemented on the device over more sophisticated techniques, such as deep learning. However, this brings its own tradeoffs. The peak detection parameters were selected based on literature of maximum resting HR. However, this biases predicted HRs towards a central value in the typical human resting HR range. 

\begin{figure}[ht]
\centering
\begin{subfigure}{.48\linewidth}
  \centering
  \includegraphics[width=1\linewidth]{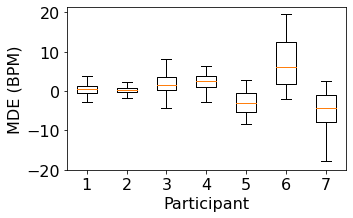}  
  \caption{MDE of aHR}
  \label{fig:AHR_MDE_parts}
\end{subfigure}
\begin{subfigure}{.48\linewidth}
  \centering
  \includegraphics[width=0.94\linewidth]{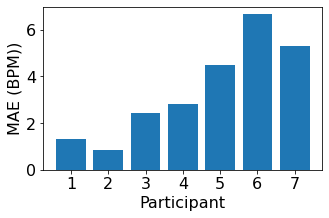}  
  \caption{MAE of aHR}
  \label{fig:AHR_MAE_parts}
\end{subfigure}
\begin{subfigure}{.48\linewidth}
  \centering
  \includegraphics[width=\linewidth]{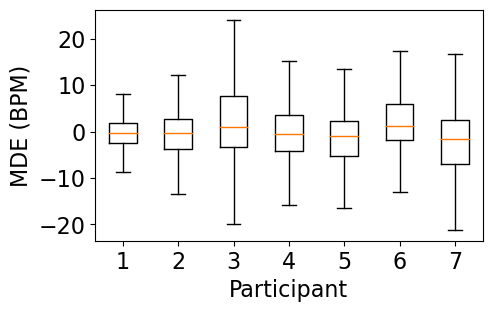}  
  \caption{MDE of iHR}
  \label{fig:IHR_MDE_parts}
\end{subfigure}
\begin{subfigure}{0.48\linewidth}
  \centering
  \includegraphics[width=0.94\linewidth]{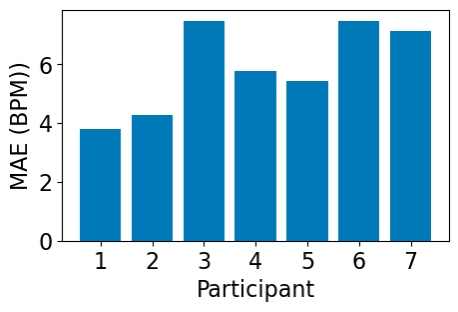}  
  \caption{MAE of iHR}
  \label{fig:IHR_MAE_parts}
\end{subfigure}
\caption{Error per participant over the dataset.}
\label{fig:participant_comp}
\end{figure}

\cref{fig:participant_comp} shows comparisons in MDE and MAE across all the participants for both aHR and iHR. It is clear that in both cases errors remain very low, however, there is a variation in error across the participants with aHR MAE ranging from 1.9\,BPM to 6.6\,BPM and iHR MAE from 3.8\,BPM to 7.6\,BPM. In particular, errors for participants 1 to 4 are low with MAE under 3\,BPM and MDE close to zero. However, errors for participants 5 to 7 are larger, with MAE reaching 7\,BPM. This is partly caused by poor adherence to the study protocols during data collection as the researchers could not meet participants in person. This lack of adherence includes not wearing devices at the correct tension or not remaining at rest through the recording period.
Additionally, other sounds in the signal collected from the abdomen, such as bowel or respiratory sounds, cause noise in the data which affects HR estimation accuracy. 

Furthermore, upon inspecting the data of the participant with the largest error (10.2\,BPM MAE), we observed that they had a significantly lower resting HR than the other participants in the study. This indicates that a more prominent, higher frequency signal was identified over the heart sounds in this participant's data, which gives rise to greater error. 
The algorithm described in \cref{sec:vitals} excludes data based on a set of rules for when there are large noise spikes or the time between heart sounds is too long. Over the whole dataset, 4.5\% of data is removed as windows contain large noise spikes. Using the retained data, HR estimations are done and 6.0\% of estimations are excluded as they exceed the resting HR range, resulting in a total missingness of \textbf{10.2\%} of the data. This is believed to be acceptable given that it makes detection feasible and prevents unrealistic heart rates from being reported.

\subsection{Longitudinal Tracking}

\begin{figure}[ht]
\begin{subfigure}{.48\linewidth}
  \centering
  \includegraphics[width=\linewidth]{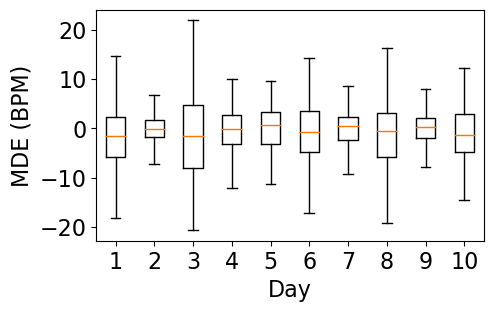}  
  \caption{MDE of iHR}
  \label{fig:MDE_day}
\end{subfigure}
\begin{subfigure}{0.48\linewidth}
  \centering
  \includegraphics[width=0.94\linewidth]{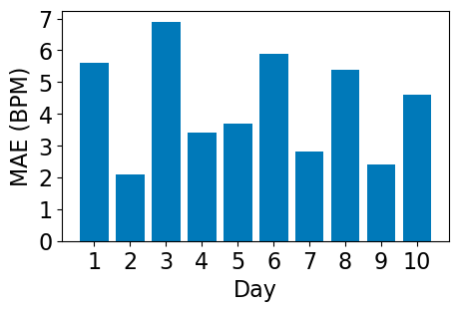}  
  \caption{MAE of iHR}
  \label{fig:MAE_day}
\end{subfigure}
\caption{Variation in iHR MDE and MAE in HR across 10 days for one participant.}
\label{fig:days}
\end{figure}

The results for HR trends across the same participant for different days is presented in \cref{fig:days}. This participant was chosen as their data had the best SNR across all the participants. The figure shows that for the same participant, error remains under 7~BPM for each day of the study. However, while remaining low, error still varies across days by up to 250\%.
This could indicate inconsistencies with wearing the device over each day or variations in the participant's resting heart rate.

\begin{figure}[ht]
    \centering
    \includegraphics[width=6.0cm]{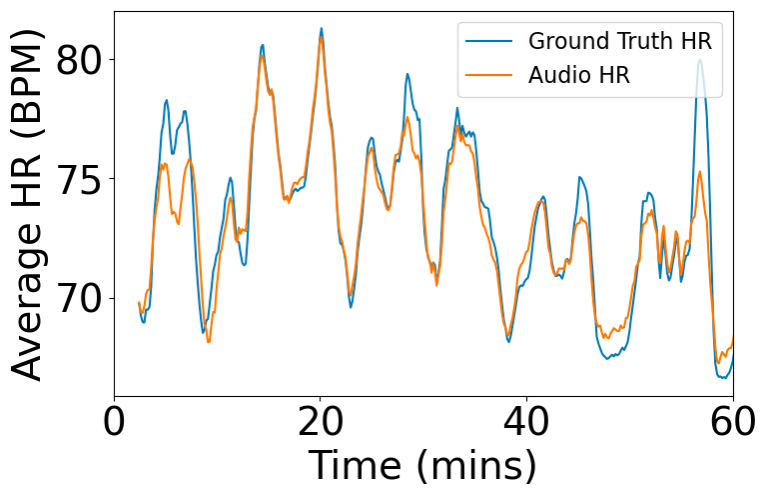}
    \caption{Tracking between the ground truth and audio estimated HR over an hour period.}
    \label{fig:track}
\end{figure}

\cref{fig:track} provides the results of HR predictions over a period of one hour. In this figure, GT HR is compared with aHR from abdominal audio. It is evident that there is a close agreement between the two measurements as the participant's HR changes over time. It can also be seen that the audio estimate underestimates at the peaks and overestimates at the troughs, consistent with \cref{fig:blandaltman}.

\subsection{Baseline Comparison}

\cref{tab:comparison} provides a baseline comparison of our results to accuracy of HRs from consumer and research grade devices as reported by Bent et al.~\cite{bent_investigating_2020}. This study assessed performance while both active and at rest, however, only the results from the rest condition are used in the comparison, since devices to monitor GI activity are designed for use at rest~\cite{wang2019flexible}. Thus, our algorithm will be suitable for most cases of ABS monitoring. It is evident from the table that our system outperforms all the consumer and research grade devices for MAE for aHR. Notably, we even outperform the best performing device, the Apple Watch, in terms of absolute error. Interestingly, although we have the lowest absolute error, our directional error is higher than the majority of the devices studied in~\cite{bent_investigating_2020}. This is due to the peak detection parameters, as previously discussed. It must be noted that these PPG devices are reporting an average HR and instantaneous HR cannot be accessed.

\begin{table}[t]
\centering
\caption{Comparison between our method and consumer and research grade devices for HR extraction at rest. The other devices results are taken from \textit{Bent et al.}~\cite{bent_investigating_2020}.}
\begin{tabular}{@{}llllll@{}}
\toprule
Method & MDE (BPM) & MAE (BPM) \\ \midrule
Apple Watch 4 & \textit{-0.09} & \textit{4.4}  \\
Fitbit Charge 2 & \textit{+0.34} & \textit{7.3}   \\
Garmin Vivosmart 3 & \textit{-0.85} & \textit{7.0}  \\
Empatica E4 & \textit{-3.9} & \textit{11.3}  \\ \midrule
ABS Average HR (Ours) & -1.2  & \textbf{3.4}  \\ \bottomrule
\end{tabular}
\label{tab:comparison}
\end{table}

%% file: sections/conclusion.tex
This work presents, for the first time, heart rate extraction (both instantaneous and average) from abdominal sounds. Creating the link between ABS and vital signs is key for an ABS measuring wearable as it allows such a wearable to obtain a holistic view of patients' general health and wellbeing, while simultaneously monitoring GI status. 

We extract average and instantaneous HR from 104 hours of data across 7 participants using lightweight signal processing techniques. Using our system, we achieve a MAE 3.4\,BPM with MDE of -1.2\,BPM for average HR, amounting to an average absolute error of 4.8\%. This feasibility study demonstrates the opportunity for integrating ABS and HR monitoring devices into the same form factor at the abdomen,  such as ordinary belts or elastic bands.  Such form factor might have competitive advantage by being fully integrated and widely accepted into existing clothing designs. 

Our system uses established signal processing techniques for HR estimation from ABS. However, for future work, segmentation with machine learning methods such as logistic regression~\cite{springer2015logistic}, or LSTMs~\cite{fernando2019heart} could be implemented to localise the heart sounds, thus enabling more accurate and consistent location of the heart sounds, thus leading to improved instantaneous HR estimations. Personalisation could be used to change the parameters of the algorithm used in this study, this could remove the affects of the bias shown in \cref{fig:blandaltman} giving higher accuracy among users with resting HR deviating from the average.  These changes could enable tracking additional vital signs, such as heart rate variability, respiratory rate, and related metrics from abdominal sounds. 